\RequirePackage{fix-cm}
\documentclass[smallextended]{svjour3}       
\smartqed  
\usepackage{graphicx}
\usepackage{float}
\usepackage{paralist}

\usepackage{amssymb}
\usepackage{bm}
\usepackage{amsmath}

\begin{document}

\title{Parallel simulation for the ultra-short laser pulses' propagation in air
}


\author{Cunliang Ma         \and
        Wenbin Lin* 
}


\institute{ 
              Institute of Electromagnetics, Southwest Jiaotong University, Chengdu 610031, China \\
              \email{wl@swjtu.edu.cn}
}

\date{Received: date / Accepted: date}

\maketitle

\begin{abstract}
A parallel 2D+1 split-step Fourier method with Crank-Nicholson scheme running on multi-core shared memory architectures is developed to study the propagation of ultra-short high-intensity laser pulses in air.
The parallel method achieves a near linear speed-up with results for the efficiency of more than $95\%$ on a 24-core machine.
This method is of great potential application in studying the long-distance propagation of the ultra-short high intensity laser pulses.
\end{abstract}

\section{Introduction}
\label{intro}
Investigation of the ultra-short high-intensity laser pulses propagation in air has been a hot topic in recent years due to its physical interest as well as its potential applications.
It is very important to predict well how the electromagnetic field of the pulse evolves as it propagates~\cite{A.Chiron1999}.
Although some analytical solutions with some approximations can be found~\cite{P.Sprangle2002,Yedierler1,Yedierler2}, in most applications the analytic approximations can not describe accurately the evolution of the pulses and we have to resort to numerical methods.
Split-step Fourier methods~\cite{Agrawal1989} with the Crank-Nicholson scheme (FCN) in the transverse direction is often employed to numerical calculate the propagation of the ultra-short laser pulses via solving the nonlinear Schr$\mathrm{\ddot{o}}$dinger equation (NLSE)\cite{P.Sprangle2002}.

Ultra-short high intensity laser pulse can convey high intensity over extended distances,
and some applications need kilometer-range calculation.
The filamentation induced by the ultra-short high-intensity pulses have been observed over several kilometers~\cite{Mechain}.
The alternative signs in the coefficients of the high-order Kerr effects allow the pulses to propagate without much energy loss in a long distance~\cite{V.Loriot2010,P.Bejot}.
However, the simulation for a long-distance propagation of the pulse is very time-consuming.
For example, it may need several weeks to calculate a kilometer-range propagation. Parallel Split-step Fourier methods~\cite{xiangming,J.Sanchez2008,Thiab2005} have been developed to solve 1D+1 NLSE. However, the 1D+1 NLSE can not be used to simulate ultra-short high intensity laser pulse propagation in air, since it is unable to describe the transverse variations of the laser pulse.

In this paper we develop a parallel FCN method to solve a 2D+1 NLSE, which can be employed to simulate ultra-short high intensity laser pulse propagation in air.
The paper is organized as follows. In section 2 we briefly introduce the NLSE which describes the ultra-short high-intensity laser pulses propagation in air.
In section 3 the serial 2D+1 FCN method for solving the NLSE equation is reviewed. Section 4 presents the parallel FCN algorithm for the 2D+1 NLSE.
Performance tests and a simulation for a long-distance pulse's propagation are given in section 5.
Conclusion is detailed in Section 6.
\section{Nonlinear propagation equation}
The wave equation for the laser electric field $\it{\vec{E}(\vec{r},t)}$ is given by~\cite{P.Sprangle2002}
\begin{eqnarray}
\it{(\nabla_{\perp}^{2}+\frac{\partial^2}{\partial{z^2}}-\frac{1}{c^2}\frac{\partial^2}{\partial{t^2}})\vec{\bm{E}}=\vec{\bm{S}}_L+\vec{\bm{S}}_{NL}}~,
\label{waveeq}
\end{eqnarray}
where $\it{\vec{\bm{S}}_L(x,y,z,t)}$, and $\it{\vec{\bm{S}}_{NL}(x,y,z,t)}$ are the linear and nonlinear source of the electric field. $\it{\vec{\bm{E}}(x,y,z,t)}$, $\it{\vec{\bm{S}}_L(x,y,z,t)}$, and $\it{\vec{\bm{S}}_{NL}(x,y,z,t)}$ can be written as~\cite{P.Sprangle2002}
\begin{subequations}
\begin{align}
\label{source}
&\it{\vec{\bm{E}}(x,y,z,t)=\frac{1}{2}A(x,y,z,t)\exp{(ik_0Z-i\omega_0t)}\hat{e}_x+c.c}~, \\
&\it{\vec{\bm{S}}_L(x,y,z,t)=\frac{1}{2}S_L(x,y,z,t)\exp{(ik_0Z-i\omega_0t)}\hat{e}_x+c.c}~, \\
&\it{\vec{\bm{S}}_{NL}(x,y,z,t)=\frac{1}{2}S_{NL}(x,y,z,t)\exp{(ik_0Z-i\omega_0t)}\hat{e}_x+c.c}~,\
\end{align}
\end{subequations}
where $c.c$ denotes the complex conjugate of the first term in the right hand side of equations. $\it{k_0}$ is the carrier wave number, $\it{\omega_0}$ is the angular frequency of the pulse, $\it{\hat{e}_x}$ denotes the unit vector in the direction of polarization.
$\it{A(x,y,z,t)}$, $\it{S_L(x,y,z,t)}$ and $\it{S_{NL}(x,y,z,t)}$ are the complex amplitudes of $\it{\vec{\bm{E}}(x,y,z,t)}$, $\it{\vec{\bm{S}}_L(x,y,z,t)}$ and $\it{\vec{\bm{S}}_{NL}(x,y,z,t)}$ respectively.

For the ultra-short and high-intensity laser pulse, $\it{S_{NL}}$ can be written as
\begin{eqnarray}
\it{S_{NL}(x,y,z,t)=S_{Kerr}+S_{Plasma}+S_{ion}}~,
\label{nlsource}
\end{eqnarray}
where $\it{S_{Kerr}}$ denotes the nonlinear contribution from bound electrons, i.e., Kerr effect,
\begin{eqnarray}
\it{S_{Kerr}(\vec{\bm{r}},t)=-\frac{2k_0}{n_0}\sum_{j}n_{2\times j}|A(\vec{\bm{r}},t)|^{2\times j}A(\vec{\bm{r}},t)}~,
\label{kerrsource}
\end{eqnarray}
here $\it{n_{2\times j}}$ is Kerr nonlinear refractive index. The nonlinear index defines a critical nonlinear self-focusing power $P_{cr}=\frac{3.77\lambda_0^2}{8\pi n_0n_2}$ for a gaussian input pulse.

The plasma source term $\it{S_{Plasma}}$ is given by
\begin{eqnarray}
\it{S_{Plasma}(\vec{\bm{r}},t)=\frac{k_0^2\omega_{pe}^2}{\omega_0}A(\vec{\bm{r}},t)}~,
\label{plasmasource}
\end{eqnarray}
where $\it{\omega_{pe}=\left(q_e^2\rho/m_e\varepsilon_0\right)^{1/2}}$ denotes the plasma frequency, $\it{q_e}$ is the plasma density generated by ionization, $\it{m_e}$ is the mass of electron and $\it{\varepsilon_0}$ is the vacuum permittivity.

The term $\it{S_{ion}}$ describes the depletion of laser energy due to ionization
\begin{eqnarray}
\it{S_{ion}(\vec{\bm{r}},t)=-ik_0\beta^{(K)}|A(\vec{\bm{r}},t)|^{2K-2}A(\vec{\bm{r}},t)}~,
\label{ionsource}
\end{eqnarray}
where $\it{\beta^{(K)}}$ is the coefficient of multiphoton ionization for the number of photons $\it{K}$.
When the wavelength is 800 nm, $\it{K}$ = 10 and $\it{\beta^{(K)}=}$ \\ $\it{\mathrm{1.27\times 10^{-126}}}
\mathrm{cm^{17}/W^9}$.~\cite{Tzortzakis2001}

Substituting Eqs. (2)-\eqref{ionsource} into Eq. \eqref{waveeq} and transforming the independent variables from $\it{z,t}$ to $\it{z,\tau}$ via
$\it{\tau=t-z/v_g}$ with $\it{v_g}$ the linear group velocity of the pulse, and applying the slowly varying envelope approximation $\it{\partial^2A/\partial z^2=0}$,
we can obtain the nonlinear schr$\ddot{o}$dinger equation describing the propagation of ultra-short and high-intensity laser pulse as follow
\begin{eqnarray}
\label{nlse}
\nonumber\it{\frac{\partial{\sl{A}}}{\partial{\sl{z}}}} &=& \it{\frac{i}{2\mathrm{k}_0}\Delta_{\perp}\sl{A}-
  \frac{ik^{''}}{2}\frac{\partial^2{\sl{A}}}{\partial{\sl{t}^2}}+\frac{ik_0}{n_0}
  \big(\sum_{j=1}^{4}n_{2*j}\left|\sl{A}\right|^{2*j}\big)\sl{A}}\\
  &&-\it{\frac{ik_0}{2}\frac{\omega_{pe}^2}{\omega_0}\sl{A}
-\frac{\beta^{(K)}}{2}\left|\sl{A}\right|^{2K-2}\sl{A}}~,
\end{eqnarray}
where $\it{\Delta_{\perp}\equiv \partial^2/\partial{r^2}+\frac{1}{r}\partial/\partial{r}}$ for cylindrically symmetric beams or $\Delta_{\perp}\equiv \partial/\partial{x^2}+\partial/\partial{y^2}$ otherwise.
In this paper the initial input pulse is chosen to have cylindrical symmetry.
$\it{k^{''}=\mathrm{0.2~fs^2/cm}}$ is the second order dispersion coefficient.

The rate equation for electron density $\it{q_e}$ can be written as
\begin{eqnarray}
\label{ionization}
\it{\frac{\partial\rho}{\partial t}=}\it{\frac{\beta^{(K)}}{K\hbar\omega_0}\left|\sl{A}\right|^{2K}\left(1-\frac{\rho}{\rho_{at}}\right)}~,
\end{eqnarray}
where $\it{\rho_{at}=\mathrm{2.7\times10^{19}~cm^{-3}}}$ is the density of the neutral atoms.

The core contribution of this work is to built a 2D+1 (time) parallel FCN solver for Eq.\eqref{nlse} and Eq.\eqref{ionization}.

\section{FCN method}

We restrict our attention to the axis-symmetric problems thus $\it{A=A(r,z,\tau)}$.
Uniform discrete lattice approximations is employed with lattice spacings $\it{\Delta r}$, $\it{\Delta z}$, and $\it{\Delta \tau}$.
$\it{A(r,z,\tau)}$ is discretized into $\it{A}$($\it{m\Delta r,n\Delta z,p \Delta \tau}$+$\\ \it{\tau_{min}}$), in which $\it{0\leqslant m < N_r\in\mathbb{N},~0\leqslant n < N_z\in\mathbb{N}}$, and $\it{0\leqslant p < N_t\in\mathbb{N}}$. In Fourier domain, $\it{\widetilde{A}(r,z,\omega)}$ can be discretized into $\it{\widetilde{A}(m\Delta r,n\Delta z,\omega_p)}$.

Let $\it{A_{m,p}^n}$ denote $\it{A}(\it{m}\it{\Delta r},\it{n}\it{\Delta z},\it{p} \it{\Delta \tau}+\it{\tau_{min}})$ and $\it{\widetilde{A}_{m,p}^n}$ denote $\it{\widetilde{A}(m\Delta r,n\Delta z,\omega_p)}$. Let $\it{A_{.,p}^n}$ represent the vector $\it{A_{(0:N_r-1),p}^n}$, $\it{A_{m,.}^n}$ represent the vector $\it{A_{m,(0:N_t-1)}^n}$, and $\it{A_{.,.}^n}$ represent the matrix $\it{A_{(0:N_r-1),(0:N_t-1)}^n}$. The NLS equation \eqref{nlse} can be written as
\begin{eqnarray}
\it{\frac{\partial{A}}{\partial{z}}=(\widehat{D}+\widehat{N})A}~,
\label{nlseq}
\end{eqnarray}
where $\widehat{D}$ and $\widehat{N}$ are the linear and nonlinear operators respectively,
\begin{eqnarray}
\label{linearpart}
\it{\widehat{D}}&=&\it{\frac{i}{2k_0}\left(\frac{\partial^2}{\partial{r^2}}+\frac{1}{r}\frac{\partial}{\partial{r}}\right)-
\frac{ik^{''}}{2}\frac{\partial^2}{\partial{t^2}},} \\
\label{nonlinearpart}
\it{\widehat{N}}&=&\it{\frac{ik_0}{n_0}\left(\sum_{j=1}^{4}n_{2*j}|A|^{2*j}\right)-\frac{ik_0}{2}\frac{
\omega_{pe}^2}{\omega_0}-\frac{\beta^{(K)}}{2}|A|^{2K-2}}.
\end{eqnarray}
In the step $n$, we first calculate the nonlinear part for a half step ($\Delta z/2$), and then calculate the linear part for a full step $\Delta z$, and finally calculate the nonlinear part for another half step ($\Delta z/2$), i.e.,
\begin{eqnarray}
\it{A^{n+1}}=\it{A^{n}} \exp({\widehat{N}\frac{\Delta z}{2}}) \exp({\widehat{D}\Delta z}) \exp({\widehat{N}\frac{\Delta z}{2}}) ,
\label{nlseq}
\end{eqnarray}

In the below we describe the algorithm in detail.

\subsection{The linear part, from $A^{n}_{.,.}$ to $A^{n+1}_{.,.}$}
The linear part is solved in the frequency domain. Firstly, $A^{n}_{.,.}$ in the time domain are transformed to $\widetilde{A^{n}_{.,.}}=\mathcal{F}(A^n_{.,.})$ in the frequency domain.
Here $\mathcal{F}$ denote Fourier transform. It follows Eq. \eqref{linearpart} that the linear-part effect satisfies
 \begin{eqnarray}
 \it{\frac{\partial{\widetilde{A}}}{\partial{z}}=\frac{i}{2k_0}\left(\frac{\partial^2}{\partial{r^2}}+
 \frac{1}{r}\frac{\partial}{\partial{r}}\right)\widetilde{A}-
 \frac{ik^{''}}{2}(i\omega)^2\widetilde{A}}~.
 \label{freqdomainlinear}
 \end{eqnarray}

 Secondly, we discretize Eq. \eqref{freqdomainlinear} using Crank-Nicholson scheme which is an implicit finite-difference method and is given by
 \begin{eqnarray}
 \label{diff1}
 \it{\frac{\partial \widetilde{A}_{m,p}^{n+\frac{1}{2}}}{\partial z}}&\approx& \it{\frac{\widetilde{A}_{m,p}^{n+1}-\widetilde{A}_{m,p}^n}{\Delta z}}~,\\
 \label{diff2}
 \it{\frac{\partial^2\widetilde{A}_{m,p}^{n+\frac{1}{2}}}{\partial r^2}}&\approx&\it{\frac{\widetilde{A}_{m+1,p}^{n}-2\widetilde{A}_{m,p}^{n}+\widetilde{A}_{m-1,p}^{n}}{2\Delta r^2}+
 \frac{\widetilde{A}_{m+1,p}^{n+1}-2\widetilde{A}_{m,p}^{n+1}+\widetilde{A}_{m-1,p}^{n+1}}{2\Delta r^2}}~,\\
 \label{diff3}
 \it{\frac{1}{r}\frac{\widetilde{A}_{m,p}^{n+\frac{1}{2}}}{\partial r}}
 &\approx&\it{\frac{\widetilde{A}_{m+1,p}^n-\widetilde{A}_{m-1,p}^n+\widetilde{A}_{m+1,p}^{n+1}-\widetilde{A}_{m-1,p}^{n+1}}{4m(\Delta r)^2}}~.
 \end{eqnarray}
Substituting Eqs. \eqref{diff1}-\eqref{diff3} into Eq. \eqref{freqdomainlinear}, and making use of the boundary conditions
\begin{eqnarray}
\label{bound}
\it{\frac{\partial{A}}{\partial{r}}|_{r=0}=0~,~~~~~~~~A|_{r=max}=0}~,
\end{eqnarray}
we can obtain a matrix equation
\begin{eqnarray}
\label{matrix}
\it{M_{abc}^p\widetilde{A}_{.,p}^{n+1}=M_{def}^p\widetilde{A}_{.,p}^n}~,
\end{eqnarray}
where
\begin{eqnarray}
 \label{matrix1}
\it{M_{abc}^p}  \equiv  \left[
 \nonumber\begin{array}{cccccc}
 \it{b_{0}} &\it{c_{0}} & & \\
 \it{a_{1}} &\it{b_{1}} &\it{c_{1}} &\\
     & \it{\ddots}& \it{\ddots} &\it{\ddots}\\
     &       &\it{a_{N_r-2}}&\it{b_{N_r-2}}&\it{c_{N_r-2}}\\
     &       &        &\it{a_{N_r-1}}  &\it{b_{N_r-1}}
 \end{array}
 \right]~,\\
\it{M_{def}^p} \equiv  \left[
 \begin{array}{cccccc}
 \it{e_{0}} &\it{f_{0}} & & \\
 \it{d_{1}} &\it{e_{1}} &\it{f_{1}} &\\
     & \it{\ddots}& \it{\ddots} &\it{\ddots}\\
     &       &\it{d_{N_r-2}}&\it{e_{N_r-2}}&\it{f_{N_r-2}}\\
     &       &        &\it{d_{N_r-1}}  &\it{e_{N_r-1}}
 \end{array}
 \right]~,
\end{eqnarray}
with 
\begin{eqnarray}
\label{MatrixIndexajp}
&&\it{a_{j}}=\it{-\frac{i\Delta z}{4k_0\Delta r^2}\left(1-\frac{1}{2j}\right)}~,~~\it{j\in[1,N_r-2]}~,\nonumber\\
\label{MatrixIndexaNrP}
&&\it{a_{N_r-1}}=\it{0},~\nonumber\\
\label{MatrixIndexb1p}
&&\it{b_{0}=1+\frac{i\Delta z}{k_0\Delta r^2}-\frac{ik^{''}\omega_p^2\Delta z}{4}}~,\nonumber\\
\label{MatrixIndexbjp}
&&\it{b_{j}=1+\frac{i\Delta z}{2k_0\Delta r^2}-\frac{ik^{''}\omega_p^2\Delta z}{4}}~,~~\it{j\in[1,N_r-2]}~,\nonumber\\
\label{MatrixIndexbNrP}
&&\it{b_{N_r-1}=1}~,\nonumber\\
\label{MatrixIndexc1p}
&&\it{c_{0}=-\frac{i\Delta z}{k_0\Delta r^2}}~,\nonumber\\
\label{MatrixIndexcjp}
&&\it{c_{j}=-\frac{i\Delta z}{4k_0\Delta r^2}\left(1-\frac{1}{2j}\right)}~,~~\it{j\in[1,N_r-2]}~,\nonumber\\
\label{MatrixIndexdjp}
&&\it{d_{j}=\frac{i\Delta z}{4k_0\Delta r^2}\left(1-\frac{1}{2j}\right)}~,~~\it{j\in[1,N_r-2]}~,\nonumber\\
\label{MatrixIndexdNrP}
&&\it{d_{N_r-1}=0}~,\nonumber\\
\label{MatrixIndexe1p}
&&\it{e_{0}=1-\frac{i\Delta z}{k_0\Delta r^2}+\frac{ik^{''}\omega_p^2\Delta z}{4}}~,\nonumber\\
\label{MatrixIndexejp}
&&\it{e_{j}=1-\frac{i\Delta z}{2k_0\Delta r^2}+\frac{ik^{''}\omega_p^2\Delta z}{4}}~,~~j\in[1,N_r-2]~,\nonumber\\
\label{MatrixIndexeNrP}
&&\it{e_{N_r-1}=0}~,\nonumber
\end{eqnarray}
\begin{eqnarray}
\label{MatrixIndexf1p}
&&\it{f_{0}=\frac{i\Delta z}{k_0\Delta r^2}}~,\nonumber\\
\label{MatrixIndexfjp}
&&\it{f_{j}=\frac{i\Delta z}{4k_0\Delta r^2}\left(1-\frac{1}{2j}\right)}~,~~\it{j\in[1,N_r-2]}~.\nonumber
\end{eqnarray}


Finally, $\it{A_{.,.}^{n+1}}$ can be obtained from $\it{\widetilde{A}_{.,.}^{n+1}}$ via inverse Fourier transformation
\begin{eqnarray}
\label{antifourier}
\it{A_{.,.}^{n+1}=\mathcal{F}^{-1}(\widetilde{A}_{.,.}^{n+1})}~.
\end{eqnarray}

The pseudo-codes for the linear-part effects of the step $n$ are listed as follows:
\begin{itemize}[]
\item first loop:$~~~~\it{m=0,...,N_r\!-\!1}$
\subitem $~~~~~$$\it{\widetilde{A}_{m,.}^n=\mathcal{F}(A_{m,.}^n)}$
\item end first loop
\item second loop:$~~~~\it{p=0,...,N_t-1}$
\subitem $~~~~~\it{\widetilde{A}^{n+1}_{.,p}=(M_{abc}^p)^{-1}M_{def}^p\widetilde{A}^n_{.,p}}$
\item end second loop
\item third loop:$~~~~{m=0,...,nr-1}$
\subitem $~~~~~$$\it{A_{m,.}^{n+1}=\mathcal{F}^{-1}(\widetilde{A}_{m,.}^{n+1})}$
\item end third loop
\end{itemize}
It is worth pointing out that the second loop involves a triangular matrix equation and thus can be solved efficiently via a chasing method.

\subsection{The nonlinear part}
The calculations of the nonlinear-part effects are divided into two stages, which can be calculated by
\begin{eqnarray}
\it{A_{.,.}^{n}\leftarrow A_{.,.}^{n}\exp{\left(\frac{\Delta z}{2}\hat{N}\right)}}~, \label{Anhalf}
\end{eqnarray}
and
\begin{eqnarray}
\it{A_{.,.}^{n+1}\leftarrow A_{.,.}^{n+1}\exp{\left(\frac{\Delta z}{2}\hat{N}\right)}}~. \label{Anhalf1}
\end{eqnarray}

The calculations of Eqs. \eqref{Anhalf} and \eqref{Anhalf1} involves the plasma density, which can be obtained via solving Eq. \eqref{ionization} with a fourth-order Runge-Kutta method. The pseudo-codes for the nonlinear-part effects in the first half step are listed as follows: 
\begin{itemize}[]
\item outer loop:$~~~~\it{m=0,...,N_r-1}$
\subitem inner loop:$~~~~\it{p=0,...,N_t-1}$
\subitem $~~~~~$calculate electron density
\subitem $~~~~~$calculate $\it{\hat{N}}$
\subitem $~~~~~$$\it{A^{n}_{m,p} \leftarrow A^n_{m,p}\exp{(\frac{\Delta z}{2}\hat{N})}}$
\subitem end inner loop
\item end outer loop
\end{itemize}

The pseudo-codes for the another half step are same as that for the first half step, and we do not repeat here.

In summary, the serial 2D+1 FCN method can be carried out by the following steps:
\begin{enumerate}
\item Input the initial data, e.g., the initial input pulse, the index of the NLS equation, the ranges of time and space, and the grid steps $\it{\Delta r, \Delta t, \Delta z}$.
\item Calculate the triangular matrix $\it{M_{abc}^p}$ and $\it{M_{def}^p}$ for the discretized frequency $\omega_p$ in Fourier domain.
\item loop:$~~~~\it{n=0,...,N_z-1}$
\subitem Calculate the first half step of the nonlinear part.
\subitem Calculate the linear part, from $\it{A_{.,.}^{n}}$ to $\it{A_{.,.}^{n+1}}$.
\subitem Calculate the second half step of the nonlinear part.

end loop
\end{enumerate}

\section{Parallel algorithm for FCN method}

Suppose we have $P$ threads to carry out the FCN method in the simulations. Set the thread's id to $1,2,3,\cdot\cdot\cdot,P$. Let $\it{R_b(i)}$ and $\it{R_e(i)}$ denote the begin grid number and the end grid number in the radial domain for the $i^{\text{th}} $ thread. Let $\it{T_b(i)}$ and $\it{T_e(i)}$ denote the begin grid number and the end grid number in the temporal domain for the $i^{\text{th}}$ thread. In order to achieve the optimal parallel efficiency, the four arrays $\it{R_b,~R_e,~T_b,}$ and $\it{T_e}$ are set as follows
\begin{eqnarray}
\it{R_b(i)}\!\!\!&=&\!\!\!\left\{
\begin{array}{rcl}
\it{(i-1) \lceil N_r/P\rceil}~, ~~~~~~~~~~~~~~~~~~~~~~~~~~~~~~~~~~~~~~~~~~~~~~~~\it{i\leqslant (N_r\%P)\!+\!1}~,\nonumber\\
\it{(N_r\%P)(\lceil N_r/P\rceil-\lfloor N_r/P\rfloor)+(i-1)\lfloor N_r/P\rfloor}~, ~~~~~~~~~~~~~~~~~~~~\text{others}~,\nonumber
\end{array}\right.\\~~\nonumber\\
\it{R_e(i)}\!\!\!&=&\!\!\!\left\{
\begin{array}{rcl}
\it{i\lceil N_r/P\rceil-1}~,\hskip 6.5cm \it{i\leqslant (N_r\%P)}~,\nonumber\\
\it{(N_r\%P)(\lceil N_r/P\rceil-\lfloor N_r/P\rfloor)+i\lfloor N_r/P\rfloor-1}~~~~~~~,\hskip 2cm \text{others}~,\nonumber
\end{array}\right.\\~~\nonumber\\
\it{T_b(i)}\!\!\!&=&\!\!\!\left\{
\begin{array}{rcl}
\it{(i-1)\lceil N_t/P\rceil}~, ~~~~~~~~~~~~~~~~~~~~~~~~~~~~~~~~~~~~~~~~~~~~~~~~~\it{i\leqslant (N_t\%P)\!+\!1}~,\nonumber\\
\it{(N_t\%P)(\lceil N_t/P\rceil-\lfloor N_t/P\rfloor)+(i-1)\lfloor N_t/P\rfloor}~, ~~~~~~~~~~~~~~~~~~~~~\text{others}~,\nonumber\\
\end{array}\right.\\~~\nonumber\\
\it{T_e(i)}\!\!\!&=&\!\!\!\left\{
\begin{array}{rcl}
\it{i\lceil N_t/P\rceil-1}~,\hskip 6.5cm \it{i\leqslant (N_t\%P)}~,\nonumber\\
\it{(N_t\%P)(\lceil N_t/P\rceil-\lfloor N_t/P\rfloor)+i\lfloor N_t/P\rfloor-1}~~~~~~~~,\hskip 2cm \text{others}~,\nonumber
\end{array}\right.
\end{eqnarray}

where $\lceil~ \rceil$ denotes the rounding-up (ceiling) operation, $\lfloor~ \rfloor$ denotes the rounding-down (floor) operation, and $\%$ denotes the modulo operation.

\begin{figure*}[h]
  \centering
  \includegraphics[width=0.6\textwidth]{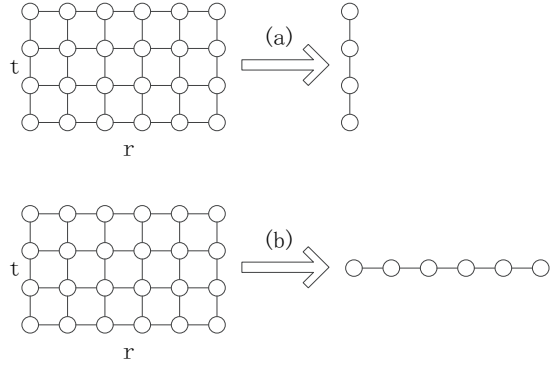}
  \caption{Data decomposition. (a) decomposition in radial domain; (b) decomposition in temporal domaim.}\label{fig1}
\end{figure*}

In the step $n$, the calculations for the discrete (reverse) Fourier transforms and the nonlinear part are decomposed in the radial domain (Fig. \ref{fig1} (a)), and the calculations for solving the linear equations \eqref{matrix} are decomposed in the temporal domain (Fig. \ref{fig1} (b)).


The pseudo-codes for the linear-part effects of the step $n$ in the $i^{\text{th}}$ thread are listed as follows:
\begin{itemize}[]
\item first loop:$~~~~\it{m=R_b(i),...,R_e(i)}$
\subitem $~~~~~$$\it{\widetilde{A}_{m,.}^n=\mathcal{F}(A_{m,.}^n)}$
\item end first loop
\item Waiting until all other P-1 threads finish the first loop
\item second loop:$~~~~\it{p=T_b(i),...,T_e(i)}$
\subitem $~~~~~\it{\widetilde{A}^{n+1}_{.,p}=(M_{abc}^p)^{-1}M_{def}^p\widetilde{A}^n_{.,p}}$
\item end second loop
\item Waiting until all other P-1 threads finish the second loop
\item third loop:$~~~~\it{m=R_b(i),...,R_e(i)}$
\subitem $~~~~~$$\it{A_{m,.}^{n+1}=\mathcal{F}^{-1}(\widetilde{A}_{m,.}^{n+1})}$
\item end third loop
\end{itemize}

The pseudo-codes for the nonlinear-part effects of the first-half step $n$ in the $i^{\text{th}}$ thread are listed as follows:
\begin{itemize}[]
\item outer loop:$~~~~\it{m=R_b(i),...,R_e(i)}$
\subitem inner loop:$~~~~\it{p=0,...,N_t-1}$
\subitem $~~~~~$calculate electron density
\subitem $~~~~~$calculate $\it{\hat{N}}$
\subitem $~~~~~$$\it{A^{n}_{m,p} \leftarrow A^n_{m,p}\exp{(\frac{\Delta z}{2}\hat{N})}}$
\subitem end inner loop
\item end outer loop
\end{itemize}

The pseudo-codes for the another half step are same as that for the first half step, and we do not repeat here.

\begin{figure*}[h]
  \centering
  \includegraphics[width=0.75\textwidth]{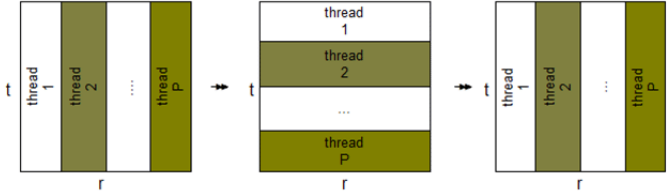}
  \caption{The re-organized procedure basing on the data decomposition.}\label{fig2}
\end{figure*}

The above parallel algorithm for the FCN method is built directly following the serial one. In order to better suite parallel programming, we re-organize the parallel algorithm into three parts basing on the data decomposition (see Fig. \ref{fig2}). The pseudo-codes for the the $n^{\text{th}}$ step in the $i^{\text{th}}$ thread are listed as follows,


\begin{itemize}[]
\item outer loop:$~~~~\it{m=R_b(i),...,R_e(i)}$
\subitem inner loop:$~~~~\it{p=0,...,N_t-1}$
\subitem $~~~~~$calculate electron density
\subitem $~~~~~$calculate $\it{\hat{N}}$
\subitem $~~~~~$$\it{A^{n}_{m,p} \leftarrow A^n_{m,p}\exp{(\frac{\Delta z}{2}\hat{N})}}$
\subitem end inner loop
\subitem $\it{\widetilde{A}_{m,.}^n=\mathcal F(A_{m,.}^n)}$
\item end outer loop
\item Waiting until all other P-1 threads finish the corresponding loop
\end{itemize}

\begin{itemize}[]
\item loop:$~~~~\it{p=T_b(i),...,T_e(i)}$
\subitem $~~~~~\it{\widetilde{A}^{n+1}_{.,p}=(M_{abc}^p)^{-1}M_{def}^p\widetilde{A}^n_{.,p}}$
\item end loop
\item Waiting until all other $P-1$ threads finish the corresponding loop
\end{itemize}

\begin{itemize}[]
\item outer loop:$~~~~\it{m=R_b(i),...,R_e(i)}$
\subitem $\it{A_{m,.}^{n+1}=\mathcal F^{-1}(\widetilde{A}_{m,.}^{n+1})}$
\subitem inner loop:$~~~~\it{p=0,...,N_t-1}$
\subitem $~~~~~$calculate electron density
\subitem $~~~~~$calculate $\it{\hat{N}}$
\subitem $~~~~~$$\it{A^{n+1}_{m,p} \leftarrow A^{n+1}_{m,p}\exp{(\frac{\Delta z}{2} \hat{N} )}}$
\subitem end inner loop
\item end outer loop
\end{itemize}

Suppose a global integer variable $Sync$ has the initial value $0$, and a global mutex $Mutex$.
The waiting other $P-1$ threads can described as follows:

\begin{itemize}[]
\item lock($Mutex$)
\subitem $Sync++$
\item unlock($Mutex$)
\item loop
\subitem if $Sync\%P$==0
\subitem $~~~~~$break
\subitem end if
\item end loop
\end{itemize}

\section{Numerical experiments}

The specifications of the computer which is used for the numerical experiments are as follows

\begin{enumerate}
\item Software Environment:
\subitem Operating System: CentOs 6.3
\subitem Development platform: g++, Pthread
\item Hardware Environment:
\subitem CPU: Intel(R) Xeo(R) E7 - 4807 @ 1.87 GHz
\subitem CPU: Cores: 24
\end{enumerate}

The initial laser pulse we consider is assumed to be a Gaussian beam~\cite{Couairon2006C}
\begin{eqnarray}
\label{gauss}
\it{A(r,0,\tau)=\sqrt{I_0}\exp{(-\frac{r^2}{r_0^2}-\frac{\tau^2}{\tau_p^2}-iC\frac{\tau^2}{\tau_p^2})}}~,
\end{eqnarray}
where $\it{r_0}$ is the beam width, $\it{\tau_p}$ is the temporal half width, $\it{I_0}$ is the input peak intensity, and $\it{C}$ denotes the chirp of the incident pulse.

\subsection{Timings and accuracy}

The performance of the parallel program is measured by Speedup, which is defined as the ratio between sequential execution time and parallel execution time~\cite{Michael2004},
\begin{eqnarray}
\label{speedup}
\mathrm{Speedup} = \frac{\mathrm{Sequential~execution~time}}{\mathrm{Parallel~execution~time}}~.
\end{eqnarray}

A numerical example ($N_r=2112,~N_t=2048$, and $N_z=1000$) is tested with different thread numbers, and Table 2 presents the comparison for the computational time with different thread numbers.
\begin{table}[h]
\centering
\begin{tabular}{|c|c|c|}
\hline
Thread number & Time (s) & Speedup  \\
\hline
1          & 5368.03 & 1                       \\
\hline
4          & 1343.821 & 3.99         \\
\hline
8          & 671.937  & 7.99         \\
\hline
12         & 450.097  & 11.93       \\
\hline
16         & 344.076  & 15.60       \\
\hline
20         & 279.084  & 19.23        \\
\hline
\end{tabular}
\caption{Computational time and speed-up ratio for different thread numbers.}
\end{table}

For the accuracy, we have checked that all the simulation results of the parallel code with different threads are the same as that of the serial code.


%
%

\subsection{One application}
\begin{figure*}[h]
  \centering
  \includegraphics[width=0.75\textwidth]{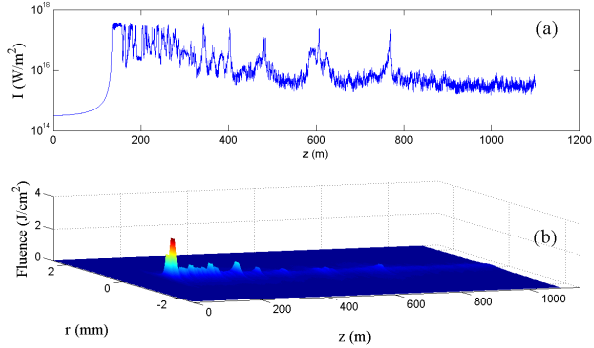}
  \caption{Propagation of a laser pulse in the atmosphere. The input beam has a Gaussian shape with
 $I_0=3.14\times 10^{14}$ W/$\mathrm{m^2}$, $\it{r_0}$ = 18 mm, $\it{\tau_0} = 300$ fs, and $\it{C} = 0$ . (a) On-axis Intensity at different propagation distance. (b) The Fluence profile as the function of the propagation distance.}\label{fig4}
\end{figure*}

We employ the parallel algorithm to simulate the propagation of the ultra-short laser pulse in air for 1.1 kilometers. In the simulation, $I_0=3.14\times 10^{14}$ W/$\mathrm{m^2}$, $\it{r_0}=18$ mm, $\it{\tau_0} = 300$ fs, and $\it{C} = 0$.
The number of grids $\it{N_r}$ = 1350, $\it{N_t}$ = 1024, and $\it{N_z}$ = $\mathrm{1.1\times{10^{6}}}$. Fig. \ref{fig4} presents the evolution of the on-axis intensity and the fluence profile of the beam. It only takes the parallel program with 20 threads less than 3 days to do the simulation, in contrast, it would require about two months for a serial code to do the same work.

\section{Conclusion}
In this paper, a parallel 2D+1 FCN method is developed which has been tested on multi-core shared memory architectures. The simulation results shows that the speed-up ratio is more than 19.2 when the thread number is 20. The parallel FCN algorithm is of great importance in the simulations for the long-distance propagation of the ultra-short laser pulse, which is very useful to many applications such as lightning control, remote sensing, and so on.

\section*{Acknowledgement}
This work was supported in part by the Ph.D. Programs Foundation of Ministry of Education of China Grant No. 20110184110016 and the National Basic Research Program of China (973 Program) Grant No. 2013CB328904, as well as the 2015 Doctoral Innovation Funds of Southwest Jiaotong University.

%
%
%


\end{document}